\documentclass{article}

\usepackage{microtype}
\usepackage{graphicx}
\usepackage{subfigure}
\usepackage{enumitem}
\usepackage{booktabs} 

\usepackage{hyperref}
\usepackage{multirow}
\usepackage{booktabs}

\usepackage[accepted]{icml2021}

\icmltitlerunning{Quantitative Evaluation of Explainable Graph Neural Networks}

\begin{document}

\twocolumn[
\icmltitle{Quantitative Evaluation of Explainable Graph Neural Networks for Molecular Property Prediction}



\icmlsetsymbol{equal}{*}

\begin{icmlauthorlist}
\icmlauthor{Jiahua Rao}{equal,to,goo}
\icmlauthor{Shuangjia Zheng}{equal,to,goo}
\icmlauthor{Yuedong Yang}{to}

\end{icmlauthorlist}

\icmlaffiliation{to}{School of Computer Science and Engineering, Sun Yat-sen University}
\icmlaffiliation{goo}{Galixir}
\icmlcorrespondingauthor{Shuangjia Zheng}{zhengshj9@mail2.sysu.edu.cn}
\icmlcorrespondingauthor{Yuedong Yang}{yangyd25@mail.sysu.edu.cn}

\icmlkeywords{Machine Learning, ICML}

\vskip 0.3in
]



\printAffiliationsAndNotice{\icmlEqualContribution} 

\begin{abstract}
Advances in machine learning have led to graph neural network-based methods for drug discovery, yielding promising results in molecular design, chemical synthesis planning, and molecular property prediction. However, current graph neural networks (GNNs) remain of limited acceptance in drug discovery is limited due to their lack of interpretability. Although this major weakness has been mitigated by the development of explainable artificial intelligence (XAI) techniques, the ``ground truth" assignment in most explainable tasks ultimately rests with subjective judgments by humans so that the quality of model interpretation is hard to evaluate in quantity. In this work, we first build three levels of benchmark datasets to quantitatively assess the interpretability of the state-of-the-art GNN models. Then we implemented recent XAI methods in combination with different GNN algorithms to highlight the benefits, limitations, and future opportunities for drug discovery. As a result, GradInput and IG generally provide the best model interpretability for GNNs, especially when combined with GraphNet and CMPNN. The integrated and developed XAI package is fully open-sourced and can be used by practitioners to train new models on other drug discovery tasks.
\end{abstract}

\section{Introduction}

Various concepts of Graph Neural Network (GNN) have been successfully adopted for drug discovery tasks in the past few years \cite{gilmer2017neural, yang2019analyzing, liu2018constrained, jin2018junction, yan2020retroxpert} such as molecular property prediction and \textit{de novo} molecular generation. Several graph neural network models have been shown to yield more promising results than the existing machine learning and quantitative structure-activity relationship (QSAR) methods for drug discovery \cite{xiong2019pushing, song2020communicative}.This advance is mostly owed to the ability of graph neural networks to effectively model molecular graph data.  
With the increasing high-quality labeled data, graph neural networks are promising to  learn better molecular representations that finally  replace decades-old hand-crafted molecular fingerprint representations.

\begin{figure*}
       \centering
	   \includegraphics[width=1\textwidth]{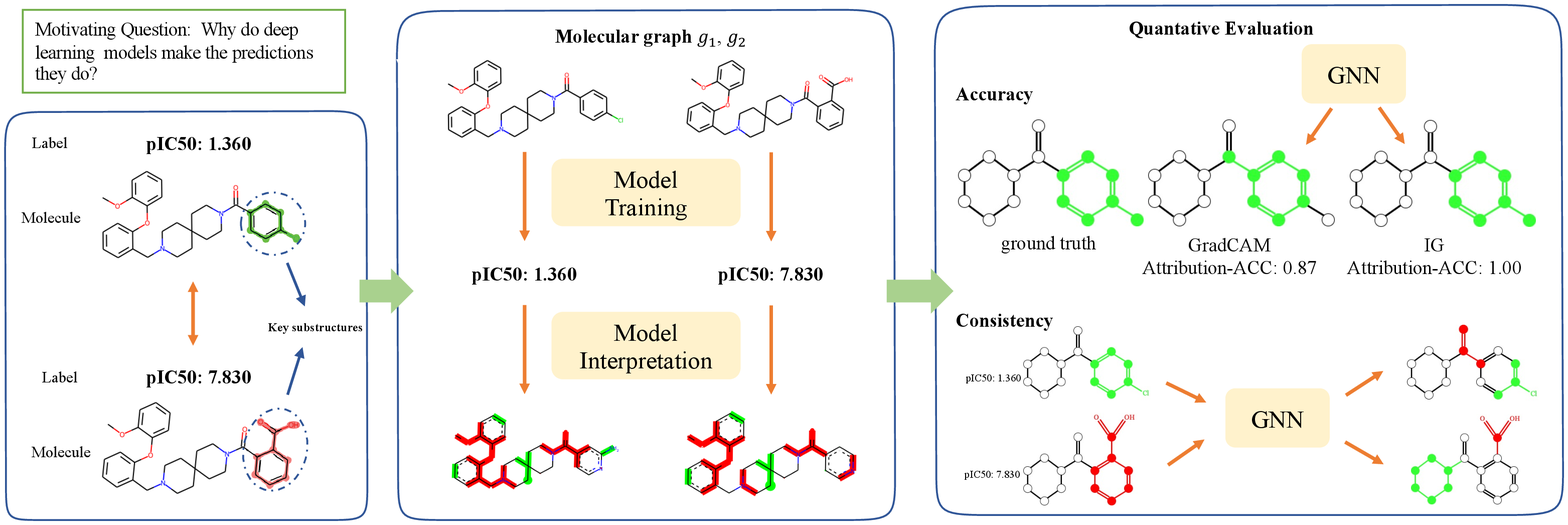}
	   \caption{Schematic of DrugXAI framework, including setups of explainable tasks  and attribution metrics. We built three levels of benchmark datasets to quantitatively assess the interpretability of GNN models. Based on the benchmarks, we implemented state-of-the-art XAI and commonly used GNN models, and provided a uniform and rigorous framework to evaluate their performances.}
\end{figure*}

Despite their promise, graph neural networks remain of limited acceptance in drug discovery partly due to their lack of interpretability, and these models are often considered as “black-box” \cite{jimenez2020drug, jimenez2021coloring}. While there have been efforts to model interpretability based on simplifications of models, feature sub-selection, or attention \cite{ribeiro2016should, sundararajan2017axiomatic, shrikumar2017learning, vaswani2017attention}, this problem is further exacerbated by the fact that these models often produce the correct answers for  wrong reasons \cite{lapuschkin2019unmasking}. Given the current pace of AI in drug discovery, there will be an increased demand for interpretability methods that help us understand and interpret the GNN models. Thus, a highly accurate and mechanically interpretable model may be the key to accelerated drug discovery with graph neural networks.

Explainable artificial intelligence (XAI) aims to help scientists to know how the model reached a particular answer and explain why the answer provided by the model is acceptable \cite{guidotti2018survey,ying2019gnnexplainer, lundberg2020local}. In drug discovery-related applications, in particular for property prediction tasks, XAI methods could help the development of graph neural networks by quantifying the molecular substructures that are critical for a given prediction and explaining how reliable a prediction is \cite{jimenez2021coloring, yu2021graph}. Feature attribution is one approach to interpretability, which measures how important the feature is to the model’s prediction of a target property. Attribution methods have been studied in the domains of drug discovery, for example, \citet{mccloskey2019using} developed an attribution method to verify whether each model trained on protein–ligand binding data learns its corresponding binding logic correctly.  \citet{jimenez2021coloring} established an Integrated Gradients (IG) feature attribution technique to examine the interpretability of GNN models in molecular property predictions. Furthermore, with the development of graph neural networks, it does not come as surprise in the fields of drug discovery to explore the interpretability of models trained with graph convolution architecture. For instance, \citet{jin2020multi} employed Monte Carlo tree search to extract molecular substructures with the help of a property predictor, which are likely responsible for each property of interest. \citet{yu2021graph} proposed a framework of Graph Information Bottleneck (GIB) for the subgraph recognition, which could recognize a compressed subgraph with minimum information loss in terms of predicting the molecular properties. These methods, which can be considered as subgraph recognition methods, are interested in directly identifying the substructures that mostly represent certain properties of molecules.

As mentioned above, many efforts have been made to mitigate the major weakness of deep learning approaches, that is the lack of causal understanding. Unfortunately, the quality and evaluation of model interpretations is hard to examine  because acquiring ground truth sub-structures or attributions requires expensive wet experiment and subjective expert judgment. Although \citet{sanchez2020evaluating} have built an open-source synthetic benchmarking suite for attribution methods on GNNs, the synthetic tasks were designed for recognize simple subgraphs such as benzene from molecules. In fact, effective XAI should reveal more complex information to scientists and render the decision-making process. For example, in toxicity prediction, there are often dozens of fragments that cause a particular toxicity, and some even have to be considered  many-to-one scenarios. Moreover, there is a kind of molecular pairs with similar structures but completely different properties, which is known as \textit{property cliffs} \cite{stumpfe2014recent}. Such realistic and sophisticated scenarios will bring more challenges and opportunities to XAI.

\begin{table*}
    \caption{Statisitics of the datasets}
    \begin{center}
    \begin{tabular}{ccccccc}
        \toprule
        Tasks & Type & Dataset Name & Compounds & Train & Test & Subgraph ground truth  \\
        \midrule
        \multirow{2}{*}{Single-rationale} & Graph classification & 3MR & 3152 & 2521 & 631 & Three-membered ring \\
        & Graph classification & Benzene & 12000 & 9600 & 2400 & Benzene ring \\
        \midrule
        \multirow{2}{*}{Multiple-rationales} & Graph classification & Mutagenicity  & 6506 & 5204 & 1302 &  Mutagenicity alerts  \\
        & Graph classification & Liver & 587 & 469 & 118 &  Hepatotoxic alerts \\
        \midrule
        \multirow{2}{*}{Property cliff} & Graph regression & hERG & 6993 & 6483 & 510 & Structural motifs  \\
        & Graph classification & CYP450 & 9122 & 9025 & 97 &  Structural motifs \\
        \bottomrule
    \end{tabular}
    \end{center}
\end{table*}

In this work, we established three levels of benchmarks, from easy to hard, for quantitatively and comprehensively assessing the interpretability of current graph neural networks and XAI modules, highlighting theirs benefits, limitations and future opportunities for drug discovery. In addition, we provide a uniform and rigorous framework to evaluate the performance of commonly-used XAI methods on several GNN model types. We examine the quality of those XAI methods through accuracy and 
consistency, which evaluate the ability of XAI methods in identifying the subgraphs that mostly represent certain properties of the molecules and perturbations that are most likely to cause \textit{property cliffs}. We offer three main contributions:
\begin{itemize}[leftmargin=*]
    \item We build three levels of benchmark datasets to quantitatively assess the interpretability of the state-of-the-art GNN models.
    \item We provide a uniform and rigorous framework to evaluate the performance of commonly-used XAI methods on several GNN model types.
    \item We provide a comprehensive overview of recent XAI methods, highlighting their benefits, limitations and future opportunities for drug discovery.
\end{itemize}

\section{Method}
In this section, We will first define the tasks and datasets used in DrugXAI and provide necessary information analyzing its effectiveness in evaluating the model interpretations. For the model interpretations, we will introduce commonly-used XAI methods on several GNN model types. Finally, detailed information about training and evaluating the XAI methods in drug discovery applications will be presented in the end.

\begin{table*}
    \caption{Prediction Performance on benchmarks}
    \begin{center}
    \begin{tabular}{cccccccc}
        \toprule
         & Dataset & 3MR & Benzene & Liver & Mutagenicity & hERG &  CYP450 \\
        \midrule
        & Metric & AUROC & AUROC & ACC & AUROC & RMSE & AUROC \\
        \midrule
        \multirow{4}{*}{GNN Model} & CMPNN & \textbf{0.998} & \textbf{0.999 }& \textbf{0.432} & \textbf{0.866} & 1.328 & \textbf{0.618} \\
        & GraphSAGE & 0.989 & \textbf{0.999} & 0.415 & 0.843 & \textbf{1.230} & 0.586 \\
        & GraphNet & 0.995 & 0.998 & 0.423 & 0.823 & 1.304 & 0.600 \\
        & GAT & 0.995 & 0.989  & 0.381 & 0.852 & 1.361 &  0.564  \\
        \bottomrule
    \end{tabular}
    \end{center}
\end{table*}

\begin{figure*}[!htbp]
       \centering
	   \includegraphics[width=1\textwidth]{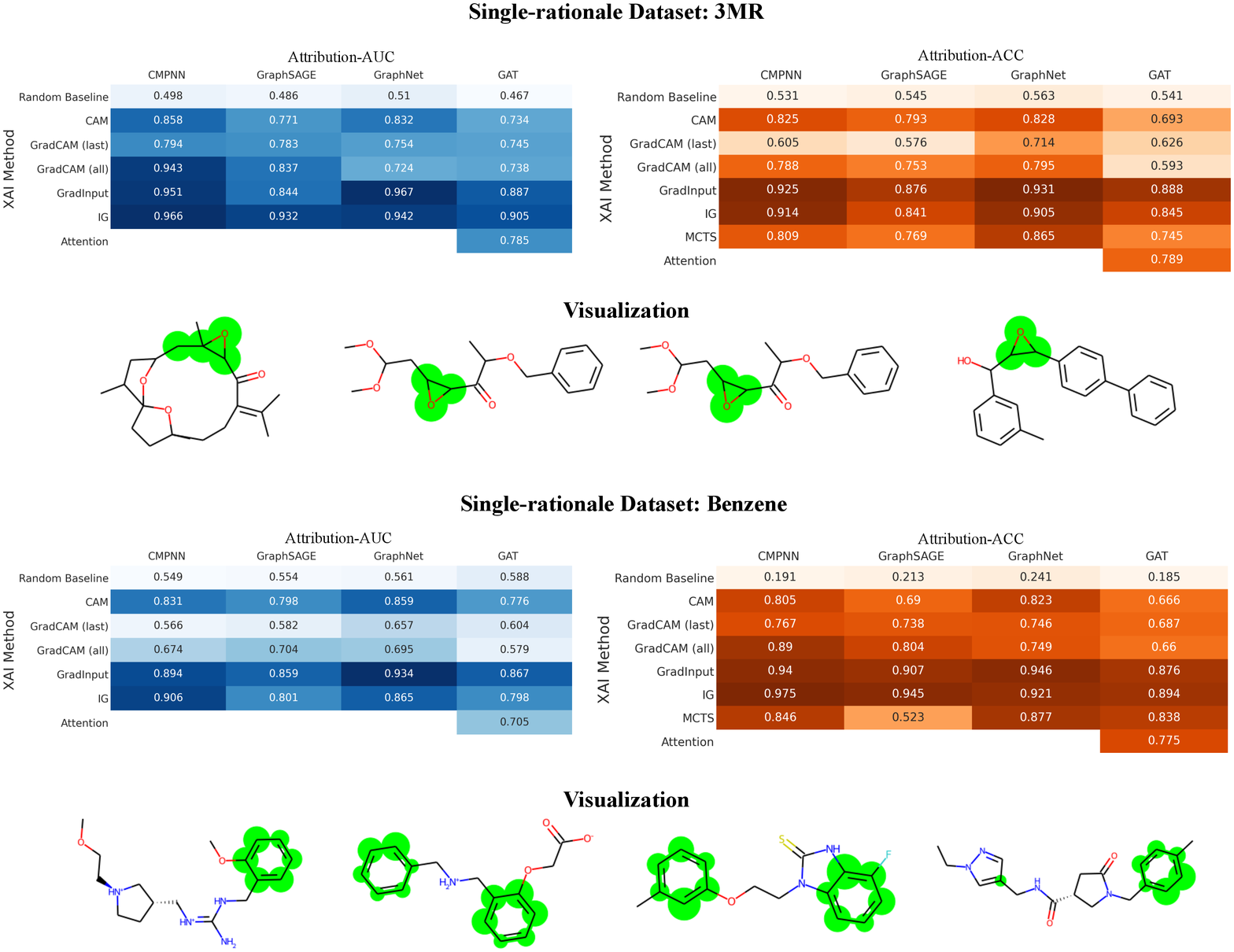}
	   \caption{\textbf{XAI method accuracy across Single-rationale tasks and model architectures. } Colors are used to distinguish two metric types — Attribution AUROC and Attribution ACC.   With the perfect predictive performances, the existing popular XAI methods on the trained GNN models are consistently superior to the random baseline and the improvements are significant.}
\end{figure*}

\subsection{Tasks and Benchmark Datasets}
Our goal is to evaluate whether current XAI methods allow GNNs to perform graph classification or regression tasks while identifying task-relevant molecular substructures  (i.e., rationales). The substructures can be a particular subgraph, the conjunction of multiple subgraphs or a local transformation between two molecular graphs (i.e, property cliff).

Correspondingly, we design three levels (single rationale, multiple rationales, and property cliff) of benchmark datasets to quantitatively assess the interpretability of the state-of-the-art GNN models. We  summarized the statistics of the datasets in Table 1.

\subsubsection{Single-rationale tasks} 
    
Following the strategy of \citet{mccloskey2019using}, we first established two single rationale benchmarks, whose goal is to identify if a molecular graph contains particular subgraphs of interest. We defined two synthetic subgraph logics: Benzene and 3MR, aiming to verify whether each model is able to capture the corresponding subgraphs correctly. In particular, we collected 50K unlabeled molecules from ZINC15 lead-like subset \cite{sterling2015zinc} and used substructure matching (Benzene and 3MR) to identify positive molecules that matched, and then randomly selected the same amount of unmatched molecules as negative samples. Both datasets are graph classification tasks, 1 for those containing a specific substructure, 0 otherwise.
   
\subsubsection{Multiple-rationales tasks}
   
In real-world scenario, there are often dozens of rationales that cause a particular property. To this end, we construct two multiple-rationales benchmarks in which ground truths are the hand-crafted structural alerts that raise the problems of hepatotoxicity and ames mutagenicity. In particular, for hepatotoxicity,  we collected the data from \citet{liu2015data}, which contains 174 hepatotoxic, 230 possible hepatotoxic and 183 non-hepatotoxic compounds, and 12 molecular rationales have been identified as structural alerts for human liver injuries. For ames mutagenicity, we used a well-established dataset compiled by \citet{hansen2009benchmark}, including 6506 compounds and corresponding ames mutagenicity results. \citet{sushko2012toxalerts} summarizes 46 structural toxic alerts that raise the mutagenicity, which have been computed as ground truth subgraphs. 


\subsubsection{Property cliff tasks}

To further explore the interpretability of the commonly-used XAI methods in extreme cases, we compiled two publicly available datasets, hERG inhibition and Cytochrome P450 Inhibition, for property cliff tasks. We use the dataset constructed by \citet{jimenez2021coloring}, which collected by a literature survey and publicly available data. Specifically, hERG inhibition is an important anti-target that must be avoided during drug development. For this endpoint, 6993 compounds with reported activity (IC50 values) in the nanomolar range were collected and IC50 values were transformed into the pIC50 scale during model training \cite{sato2018construction}. Inhibition of cytochrome P450 (CYP450) enzymes is the most common mechanism leading to drug–drug interactions. For this endpoint, 9120 CYP3A4 inhibitors and substrates with binary activity information (active/inactive) were determined by  \citet{veith2009comprehensive}.

Previously,  \citet{jimenez2021coloring} established these endpoint datasets but missed ground truths for interpretations. The “ground truth” in their experiments ultimately rests with subjective human judgment. Differently, we have constructed a quantitatively assessable scheme based on the characteristics of the property cliff.  Particularly, we first split these datasets into training and testing, forcing those eligible property cliff molecular pairs to appear only in the test set. Then, the property cliff pairs of molecules with similar structure but completely different properties were captured by MMPA technique \cite{hussain2010computationally} in the RDKit package and the ground truth attributions were computed according to the local differences between the molecular pairs. 
For the property cliff dataset which is the graph regression task,  we define a local transformation to create a pair of active cliff pairs when it causes more than a 100-fold change ($\bigtriangleup$ pIC50 $\textgreater$ 2) in properties.



\subsection{Graph Neural Network}
When viewing molecules as graphs with atoms as nodes and chemical bonds as edges, Graph Neural Network (GNN) \cite{kipf2017semi,gilmer2017neural} is a neural network that takes a graph as input and outputs a graph with the same topology, but with updated node, edge and/or graph-level information. One key feature of the GNNs we study is the message passing function, which allows nodes to update their states by aggregating feature information from neighboring nodes and edges. Depending on the message passing strategy, the message can contain information about the node, edge, or the global information. 

Our experiments use four existing popular GNN architectures distinguished by their message-passing strategies. The first is a framework for inductive representation learning (GraphSAGE)\cite{hamilton2017inductive}, which is used to generate low-dimensional vector representations for nodes. The second model is the Communicative Message Passing Neural Network (CMPNN) \cite{song2020communicative} which improve the molecular embedding by strengthening the message interactions between nodes and edges through a communicative kernel. Another model is the Graph Attention Network (GAT)\cite{velickovic2018graph} which aggregates node information via an attention mechanism. The fourth is Graph Nets\cite{battaglia2018relational} in which the message mechanism utilizes a global features vector in addition to node and edge features.

\begin{figure*}[!htbp]
       \centering
	   \includegraphics[width=1\textwidth]{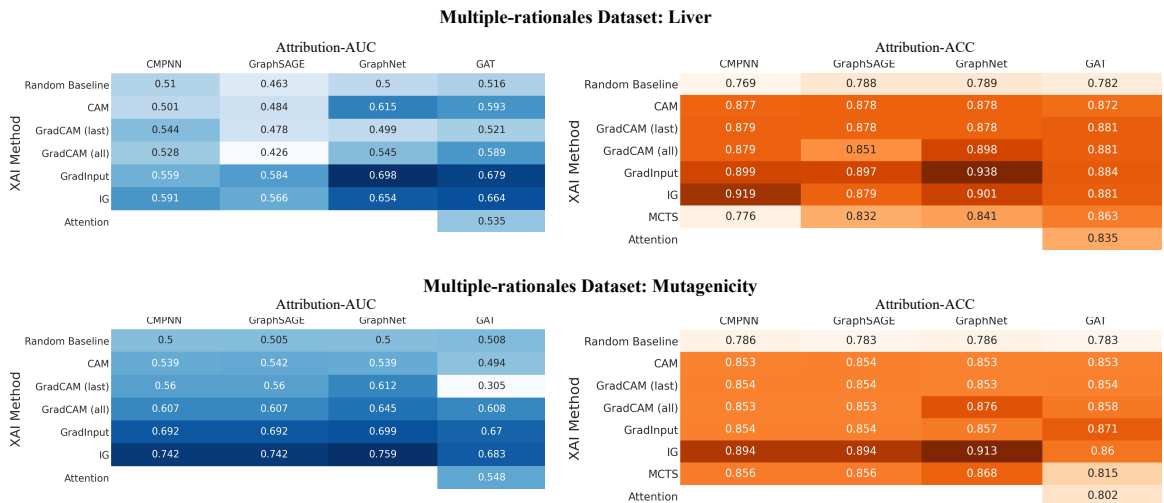}
	   \caption{\textbf{XAI method accuracy across Multiple-rationales tasks and model architectures. } Colors are used to distinguish two metric types — Attribution AUROC and Attribution ACC.  GradInput and IG consistently well across these tasks and GNN models. The attribution performances, for example, the Attribution-ACC on Liver datasets ranging between 0.835 and 0.938, have shown that the GNN models with XAI methods were able to detect the implicit relationships compared to the random baselines.}
\end{figure*}

\subsection{XAI methods}

An XAI method $A$ takes a model $M$ and a molecular graph $G$ as inputs to generate an attribution score $G_A = (v_A, e_A) $ where $v_A$ and $e_A$ are node and edge weightings relevant for predicting property $y$. These weightings can be visualized as a heatmap superimposed on a graph. Our ground-truths for attributions are node-level, so we redistribute edge attributions equally onto their endpoint nodes’ attributions. In our framework, we utilize the following methods for molecular graphs: GradInput \cite{shrikumar2017learning}, GradCAM \cite{selvaraju2017grad}, Integrated Gradients \cite{sundararajan2017axiomatic}, CAM \cite{zhou2016learning}, MCTS \cite{jin2020multi} and Attention weights \cite{vaswani2017attention}. 

Class Activation Map (CAM) is a feature attribution method, which uses a global average pooling (GAP) layer prior to class outputs to obtain attributions. CAM attributions could be derived by multiplying the final convolutional layer’s feature map activations with the output weights at the last message passing layer. GradInput attributions are calculated by the element-wise product of the input graph with the gradient of property with respect to the input node and edge features while GradCAM extends GradInput by using intermediate activations in which corresponds to the element-wise product of the activations of intermediate message-passing layer with the gradient of property. Integrated Gradients (IG) integrates the element-wise product of an interpolated input with the gradient of property. Attention mechanism method is specific to the GAT model, which will produce attention scores on edges to adjacent nodes. We can use these attention scores as a measure of importance for propagating information relevant to the predictive task. Monte Carlo Tree Search (MCTS), a subgraph recognition method, is also used to extract candidate rationales from molecules with the help of a property predictor.

\section{Experiments}
In this section, We first present the predictive performance evaluation for graph classification and regression tasks on different benchmarks. It's necessary as the interpretations generated by the model with poor predictive capability bear little trust. Next, we evaluate the attribution performance of commonly-used XAI methods on the trained GNN models. Finally, we analyze the performances on the three levels of benchmarks, from easy to hard, from recognizing particular subgraphs of interest to identifying property cliffs. All the GNN models and XAI methods were implemented based on \citet{rao2021molrep} and \citet{sanchez2020evaluating}.  Particularly, no explicit hyper-parameter optimization is performed on any model training to avoid unnecessary model selection bias. All the implementation details can be found in \url{https://github.com/biomed-AI/MolRep}.

\subsection{Predictive Performance}
The results of predictive performance on our benchmarks are presented in Table 2, where the root-mean-square error (RMSE) of regression tasks, the receiver-operator characteristic area under the curve (AUROC) of classification tasks and the accuracy (ACC) of multi-classification tasks are reported. We observed that all trained models showed perfect predictive capabilities on the single-rationale datasets, with AUROC ranging between 0.989 and 0.999, as shown in Table 2. These values suggested that the meaningful molecular graph features were identified by all trained models in the learning process. On the multiple-rationales benchmarks, the results obtained were also markedly better than random, with ACC ranging between 0.381 and 0.432 for the three-classification dataset, and AUROC ranging between 0.823 and 0.866 for the binary classification model. Finally, on the most difficult datasets, the poorest performances of the existing GNN models were obtained, suggesting that current GNN models have difficulties to capture the property-relevant features during the learning process.

\subsection{Single Rationale Recognition}

On the single-rationale datasets, we used Attribution-AUC and Attribution-ACC, designed by \citet{mccloskey2019using}, to assess attribution accuracy and consistency in identifying the corresponding substructure correctly. If multiple attributions are valid (e.g., a subgraph is present twice in a graph), we take the maximum attribution value of all possible solutions. Particularly, the subgraph recognition methods such as MCTS could not be evaluated by Attribution-AUC since their outputs are directly subgraphs rather than a node probability. 

Jointly analyzing Table 2 and figure 2, we found that with the perfect predictive performances, the existing popular XAI methods on the trained GNN models are consistently superior to the random baseline and the improvements are significant, suggesting that the trained models have identified the meaningful features that are relevant to molecular property and the current XAI approaches could effectively recognize subgraph logics corresponding. This makes sense since the graph labels of single-rationale datasets are whether the corresponding molecule contains the particular subgraphs of interest. The direct relationship between target rationales and graph labels helps the models to capture accurate information.

\begin{figure*}
       \centering
	   \includegraphics[width=1\textwidth]{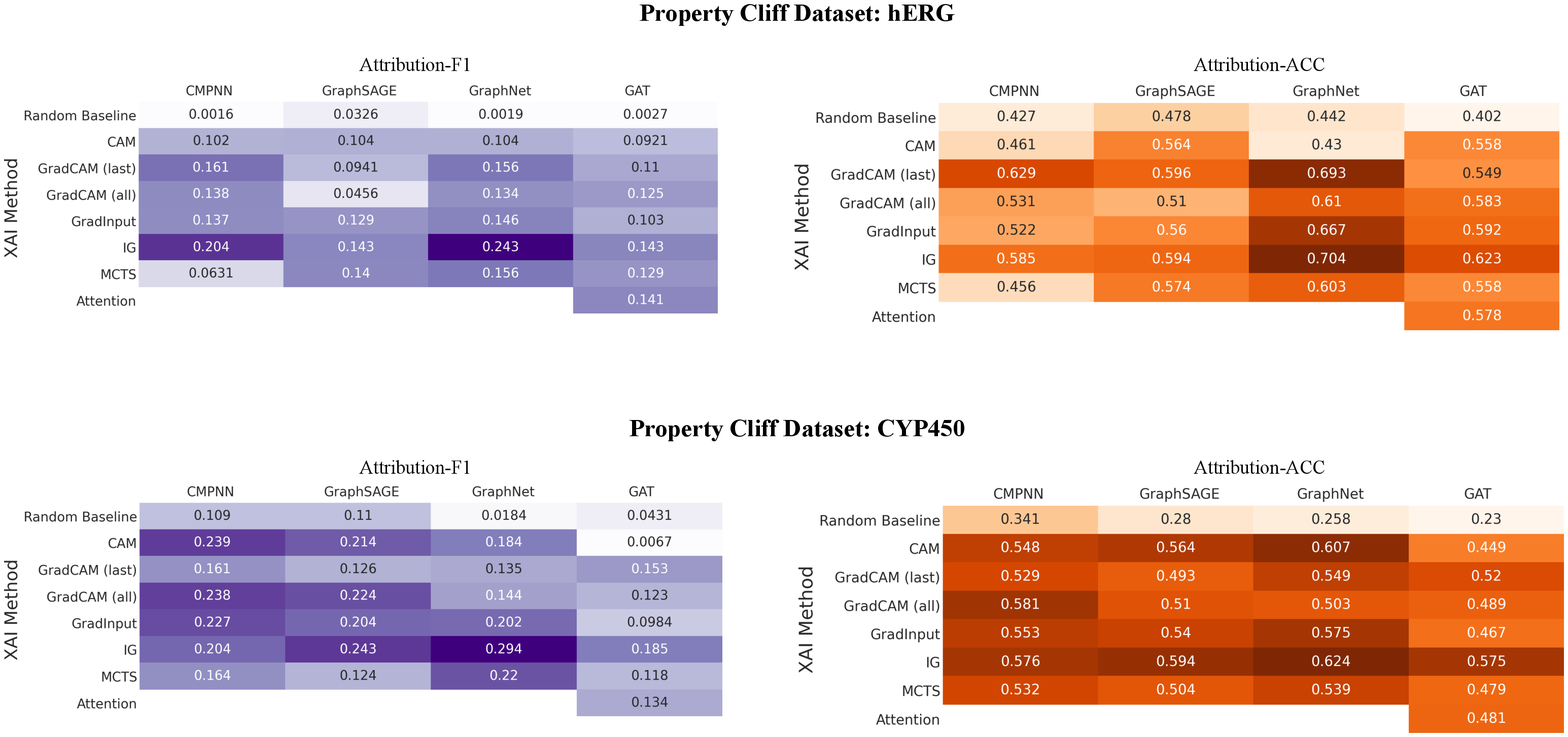}
	   \caption{\textbf{XAI method accuracy across Property-Cliff tasks and model architectures. } Colors are used to distinguish two metric types — Attribution F1 and Attribution ACC.  Most of the XAI methods failed to recognize property cliffs on the Property-Cliff tasks. Among them, IG and GradInput performed relatively well. }
\end{figure*}

\subsection{Multiple Rationales Recognition}
Furthermore, in real-world scenario, the relationships between the molecular structures and their related properties are often implicit. Therefore, whether the current GNN model with the popular XAI methods is capable of capturing the implicit relationships is the motivation for us to conduct the attribution experiments on multiple-rationales datasets.

As shown in Figure 3 and Table 2, owing to the prediction performances have demonstrated that the useful graph features were identified during the learning process, the attribution performances, for example, the Attribution-ACC on Liver datasets ranging between 0.835 and 0.938, have shown that the GNN models with XAI methods were able to detect the implicit relationships compared to the random baselines. More specifically, we could observe that GraphNet and CMPNN with high predictive performance have better attribution performance than other baselines across the popular XAI methods. And GradInput and IG across tasks and model architectures have better attribution performances. We attribute the improvements to the utilization of the gradients of the predicted property, which directly provide detailed insight into the relationship between the input features and the property.

\subsection{Property Cliff Identification}

Finally, to explore the interpretability of GNN models on the more complex relationships between substructures and graph properties, we evaluated the capabilities of the models to recognize property cliffs. Since the ground truths (local transformations) on these datasets have been classified to positive and negative based on the change of property, we only report Attribution-ACC and Attribution-F1 to assess the attribution accuracy and verify whether each model learns the corresponding key substructure transformations relevant for predicting property correctly. 

Jointly Analyzing Figure 4 and Table 2, we observed that most of the XAI methods failed to recognize property cliffs effectively under this condition, with low Attribution-F1 ranged from 0.104 to 0.294 on hERG and CYP datasets. There are two possible reasons for the results. Firstly, the poor predictive performances of the existing GNN models were obtained, suggesting that current GNN models were unable to capture the property-relevant features on the property-cliff tasks. Secondly, the relationships between property cliffs and their corresponding property are extremely diverse and complex. But considered from another perspective, on property cliff benchmarks, GNN methods are ten or even tens of times better than the random baseline, proving that although the problem is very difficult, the GNNs have captured some key features from it.

\section{Discussion and conclusions}
In this work we have established a series of XAI benchmarks in the context of drug discovery and created a framework to quantitatively evaluate attribution methods in GNNs. We expect that our benchmark and framework will aid in developing better methodology for interpretability in molecular graph tasks. Overall, our experiments show existing popular XAI methods have good performance with GNN architectures in recognizing regular subgraphs in molecules but performed poorly in identifying the property cliff on realistic datasets. As a final caveat, we find that XAI techniques that are directly related to predicted labels like GradInput and IG also tend to have better attribution performance, across many models. This also hints that current XAI methods cannot yet be used as a recipe for drug discovery, requiring significant human expertise for correct interpretation. The reasons for these phenomena remain a topic of further study.

There also remains much room for improvement in interpretation performance for GNNs. Nonetheless, further development of XAI applications would greatly benefit from this meaningful benchmarking. We hope that our quantitative benchmark and framework of interpretations in GNNs are useful for developing new XAI methods. Our framework and dataset are available for further study, broadening the development of explainable artificial intelligence in drug discovery-related applications.

\nocite{langley00}

\bibliography{main}

\begin{thebibliography}{37}
\providecommand{\natexlab}[1]{#1}
\providecommand{\url}[1]{\texttt{#1}}
\expandafter\ifx\csname urlstyle\endcsname\relax
  \providecommand{\doi}[1]{doi: #1}\else
  \providecommand{\doi}{doi: \begingroup \urlstyle{rm}\Url}\fi

\bibitem[Battaglia et~al.(2018)Battaglia, Hamrick, Bapst, Sanchez-Gonzalez,
  Zambaldi, Malinowski, Tacchetti, Raposo, Santoro, Faulkner,
  et~al.]{battaglia2018relational}
Battaglia, P.~W., Hamrick, J.~B., Bapst, V., Sanchez-Gonzalez, A., Zambaldi,
  V., Malinowski, M., Tacchetti, A., Raposo, D., Santoro, A., Faulkner, R.,
  et~al.
\newblock Relational inductive biases, deep learning, and graph networks.
\newblock \emph{arXiv preprint arXiv:1806.01261}, 2018.

\bibitem[Gilmer et~al.(2017)Gilmer, Schoenholz, Riley, Vinyals, and
  Dahl]{gilmer2017neural}
Gilmer, J., Schoenholz, S.~S., Riley, P.~F., Vinyals, O., and Dahl, G.~E.
\newblock Neural message passing for quantum chemistry.
\newblock In \emph{International Conference on Machine Learning}, pp.\
  1263--1272. PMLR, 2017.

\bibitem[Guidotti et~al.(2018)Guidotti, Monreale, Ruggieri, Turini, Giannotti,
  and Pedreschi]{guidotti2018survey}
Guidotti, R., Monreale, A., Ruggieri, S., Turini, F., Giannotti, F., and
  Pedreschi, D.
\newblock A survey of methods for explaining black box models.
\newblock \emph{ACM computing surveys (CSUR)}, 51\penalty0 (5):\penalty0 1--42,
  2018.

\bibitem[Hamilton et~al.(2017)Hamilton, Ying, and
  Leskovec]{hamilton2017inductive}
Hamilton, W.~L., Ying, R., and Leskovec, J.
\newblock Inductive representation learning on large graphs.
\newblock \emph{arXiv preprint arXiv:1706.02216}, 2017.

\bibitem[Hansen et~al.(2009)Hansen, Mika, Schroeter, Sutter, Ter~Laak,
  Steger-Hartmann, Heinrich, and M{\"{u}}ller]{hansen2009benchmark}
Hansen, K., Mika, S., Schroeter, T., Sutter, A., Ter~Laak, A., Steger-Hartmann,
  T., Heinrich, N., and M{\"{u}}ller, K.-R.
\newblock Benchmark data set for in silico prediction of ames mutagenicity.
\newblock \emph{Journal of chemical information and modeling}, 49\penalty0
  (9):\penalty0 2077--2081, 2009.

\bibitem[Hussain \& Rea(2010)Hussain and Rea]{hussain2010computationally}
Hussain, J. and Rea, C.
\newblock Computationally efficient algorithm to identify matched molecular
  pairs (mmps) in large data sets.
\newblock \emph{Journal of chemical information and modeling}, 50\penalty0
  (3):\penalty0 339--348, 2010.

\bibitem[Jim{\'e}nez-Luna et~al.(2020)Jim{\'e}nez-Luna, Grisoni, and
  Schneider]{jimenez2020drug}
Jim{\'e}nez-Luna, J., Grisoni, F., and Schneider, G.
\newblock Drug discovery with explainable artificial intelligence.
\newblock \emph{Nature Machine Intelligence}, 2\penalty0 (10):\penalty0
  573--584, 2020.

\bibitem[Jim{\'e}nez-Luna et~al.(2021)Jim{\'e}nez-Luna, Skalic, Weskamp, and
  Schneider]{jimenez2021coloring}
Jim{\'e}nez-Luna, J., Skalic, M., Weskamp, N., and Schneider, G.
\newblock Coloring molecules with explainable artificial intelligence for
  preclinical relevance assessment.
\newblock \emph{Journal of Chemical Information and Modeling}, 61\penalty0
  (3):\penalty0 1083--1094, 2021.

\bibitem[Jin et~al.(2018)Jin, Barzilay, and Jaakkola]{jin2018junction}
Jin, W., Barzilay, R., and Jaakkola, T.
\newblock Junction tree variational autoencoder for molecular graph generation.
\newblock In \emph{International Conference on Machine Learning}, pp.\
  2323--2332. PMLR, 2018.

\bibitem[Jin et~al.(2020)Jin, Barzilay, and Jaakkola]{jin2020multi}
Jin, W., Barzilay, R., and Jaakkola, T.
\newblock Multi-objective molecule generation using interpretable
  substructures.
\newblock In \emph{International Conference on Machine Learning}, pp.\
  4849--4859. PMLR, 2020.

\bibitem[Kipf \& Welling(2017)Kipf and Welling]{kipf2017semi}
Kipf, T.~N. and Welling, M.
\newblock Semi-supervised classification with graph convolutional networks.
\newblock In \emph{International Conference on Learning Representations
  (ICLR)}, 2017.

\bibitem[Langley(2000)]{langley00}
Langley, P.
\newblock Crafting papers on machine learning.
\newblock In Langley, P. (ed.), \emph{Proceedings of the 17th International
  Conference on Machine Learning (ICML 2000)}, pp.\  1207--1216, Stanford, CA,
  2000. Morgan Kaufmann.

\bibitem[Lapuschkin et~al.(2019)Lapuschkin, W{\"a}ldchen, Binder, Montavon,
  Samek, and M{\"u}ller]{lapuschkin2019unmasking}
Lapuschkin, S., W{\"a}ldchen, S., Binder, A., Montavon, G., Samek, W., and
  M{\"u}ller, K.-R.
\newblock Unmasking clever hans predictors and assessing what machines really
  learn.
\newblock \emph{Nature communications}, 10\penalty0 (1):\penalty0 1--8, 2019.

\bibitem[Liu et~al.(2018)Liu, Allamanis, Brockschmidt, and
  Gaunt]{liu2018constrained}
Liu, Q., Allamanis, M., Brockschmidt, M., and Gaunt, A.~L.
\newblock Constrained graph variational autoencoders for molecule design.
\newblock \emph{arXiv preprint arXiv:1805.09076}, 2018.

\bibitem[Liu et~al.(2015)Liu, Yu, and Wallqvist]{liu2015data}
Liu, R., Yu, X., and Wallqvist, A.
\newblock Data-driven identification of structural alerts for mitigating the
  risk of drug-induced human liver injuries.
\newblock \emph{Journal of cheminformatics}, 7\penalty0 (1):\penalty0 1--8,
  2015.

\bibitem[Lundberg et~al.(2020)Lundberg, Erion, Chen, DeGrave, Prutkin, Nair,
  Katz, Himmelfarb, Bansal, and Lee]{lundberg2020local}
Lundberg, S.~M., Erion, G., Chen, H., DeGrave, A., Prutkin, J.~M., Nair, B.,
  Katz, R., Himmelfarb, J., Bansal, N., and Lee, S.-I.
\newblock From local explanations to global understanding with explainable ai
  for trees.
\newblock \emph{Nature machine intelligence}, 2\penalty0 (1):\penalty0 56--67,
  2020.

\bibitem[McCloskey et~al.(2019)McCloskey, Taly, Monti, Brenner, and
  Colwell]{mccloskey2019using}
McCloskey, K., Taly, A., Monti, F., Brenner, M.~P., and Colwell, L.~J.
\newblock Using attribution to decode binding mechanism in neural network
  models for chemistry.
\newblock \emph{Proceedings of the National Academy of Sciences}, 116\penalty0
  (24):\penalty0 11624--11629, 2019.

\bibitem[Rao et~al.(2021)Rao, Zheng, Song, Chen, Li, Xie, Yang, Chen, and
  Yang]{rao2021molrep}
Rao, J., Zheng, S., Song, Y., Chen, J., Li, C., Xie, J., Yang, H., Chen, H.,
  and Yang, Y.
\newblock Molrep: A deep representation learning library for molecular property
  prediction.
\newblock \emph{bioRxiv}, 2021.

\bibitem[Ribeiro et~al.(2016)Ribeiro, Singh, and Guestrin]{ribeiro2016should}
Ribeiro, M.~T., Singh, S., and Guestrin, C.
\newblock " why should i trust you?" explaining the predictions of any
  classifier.
\newblock In \emph{Proceedings of the 22nd ACM SIGKDD international conference
  on knowledge discovery and data mining}, pp.\  1135--1144, 2016.

\bibitem[Sanchez-Lengeling et~al.(2020)Sanchez-Lengeling, Wei, Lee, Reif, Wang,
  Qian, McCloskey, Colwell, and Wiltschko]{sanchez2020evaluating}
Sanchez-Lengeling, B., Wei, J., Lee, B., Reif, E., Wang, P., Qian, W.~W.,
  McCloskey, K., Colwell, L., and Wiltschko, A.
\newblock Evaluating attribution for graph neural networks.
\newblock \emph{Advances in Neural Information Processing Systems}, 33, 2020.

\bibitem[Sato et~al.(2018)Sato, Yuki, Ogura, and Honma]{sato2018construction}
Sato, T., Yuki, H., Ogura, K., and Honma, T.
\newblock Construction of an integrated database for herg blocking small
  molecules.
\newblock \emph{PLoS One}, 13\penalty0 (7):\penalty0 e0199348, 2018.

\bibitem[Selvaraju et~al.(2017)Selvaraju, Cogswell, Das, Vedantam, Parikh, and
  Batra]{selvaraju2017grad}
Selvaraju, R.~R., Cogswell, M., Das, A., Vedantam, R., Parikh, D., and Batra,
  D.
\newblock Grad-cam: Visual explanations from deep networks via gradient-based
  localization.
\newblock In \emph{Proceedings of the IEEE international conference on computer
  vision}, pp.\  618--626, 2017.

\bibitem[Shrikumar et~al.(2017)Shrikumar, Greenside, and
  Kundaje]{shrikumar2017learning}
Shrikumar, A., Greenside, P., and Kundaje, A.
\newblock Learning important features through propagating activation
  differences.
\newblock In \emph{International Conference on Machine Learning}, pp.\
  3145--3153. PMLR, 2017.

\bibitem[Song et~al.(2020)Song, Zheng, Niu, Fu, Lu, and
  Yang]{song2020communicative}
Song, Y., Zheng, S., Niu, Z., Fu, Z.-H., Lu, Y., and Yang, Y.
\newblock Communicative representation learning on attributed molecular graphs.
\newblock In \emph{IJCAI}, 2020.

\bibitem[Sterling \& Irwin(2015)Sterling and Irwin]{sterling2015zinc}
Sterling, T. and Irwin, J.~J.
\newblock Zinc 15--ligand discovery for everyone.
\newblock \emph{Journal of chemical information and modeling}, 55\penalty0
  (11):\penalty0 2324--2337, 2015.

\bibitem[Stumpfe et~al.(2014)Stumpfe, Hu, Dimova, and
  Bajorath]{stumpfe2014recent}
Stumpfe, D., Hu, Y., Dimova, D., and Bajorath, J.
\newblock Recent progress in understanding activity cliffs and their utility in
  medicinal chemistry: miniperspective.
\newblock \emph{Journal of medicinal chemistry}, 57\penalty0 (1):\penalty0
  18--28, 2014.

\bibitem[Sundararajan et~al.(2017)Sundararajan, Taly, and
  Yan]{sundararajan2017axiomatic}
Sundararajan, M., Taly, A., and Yan, Q.
\newblock Axiomatic attribution for deep networks.
\newblock In \emph{International Conference on Machine Learning}, pp.\
  3319--3328. PMLR, 2017.

\bibitem[Sushko et~al.(2012)Sushko, Salmina, Potemkin, Poda, and
  Tetko]{sushko2012toxalerts}
Sushko, I., Salmina, E., Potemkin, V.~A., Poda, G., and Tetko, I.~V.
\newblock Toxalerts: a web server of structural alerts for toxic chemicals and
  compounds with potential adverse reactions, 2012.

\bibitem[Vaswani et~al.(2017)Vaswani, Shazeer, Parmar, Uszkoreit, Jones, Gomez,
  Kaiser, and Polosukhin]{vaswani2017attention}
Vaswani, A., Shazeer, N., Parmar, N., Uszkoreit, J., Jones, L., Gomez, A.~N.,
  Kaiser, L., and Polosukhin, I.
\newblock Attention is all you need.
\newblock \emph{arXiv preprint arXiv:1706.03762}, 2017.

\bibitem[Veith et~al.(2009)Veith, Southall, Huang, James, Fayne, Artemenko,
  Shen, Inglese, Austin, Lloyd, et~al.]{veith2009comprehensive}
Veith, H., Southall, N., Huang, R., James, T., Fayne, D., Artemenko, N., Shen,
  M., Inglese, J., Austin, C.~P., Lloyd, D.~G., et~al.
\newblock Comprehensive characterization of cytochrome p450 isozyme selectivity
  across chemical libraries.
\newblock \emph{Nature biotechnology}, 27\penalty0 (11):\penalty0 1050--1055,
  2009.

\bibitem[Veli{\v{c}}kovi{\'{c}} et~al.(2018)Veli{\v{c}}kovi{\'{c}}, Cucurull,
  Casanova, Romero, Li{\`{o}}, and Bengio]{velickovic2018graph}
Veli{\v{c}}kovi{\'{c}}, P., Cucurull, G., Casanova, A., Romero, A., Li{\`{o}},
  P., and Bengio, Y.
\newblock {Graph Attention Networks}.
\newblock \emph{International Conference on Learning Representations}, 2018.
\newblock URL \url{https://openreview.net/forum?id=rJXMpikCZ}.
\newblock accepted as poster.

\bibitem[Xiong et~al.(2019)Xiong, Wang, Liu, Zhong, Wan, Li, Li, Luo, Chen,
  Jiang, et~al.]{xiong2019pushing}
Xiong, Z., Wang, D., Liu, X., Zhong, F., Wan, X., Li, X., Li, Z., Luo, X.,
  Chen, K., Jiang, H., et~al.
\newblock Pushing the boundaries of molecular representation for drug discovery
  with the graph attention mechanism.
\newblock \emph{Journal of medicinal chemistry}, 63\penalty0 (16):\penalty0
  8749--8760, 2019.

\bibitem[Yan et~al.(2020)Yan, Ding, Zhao, Zheng, Yang, Yu, and
  Huang]{yan2020retroxpert}
Yan, C., Ding, Q., Zhao, P., Zheng, S., Yang, J., Yu, Y., and Huang, J.
\newblock Retroxpert: Decompose retrosynthesis prediction like a chemist.
\newblock \emph{arXiv preprint arXiv:2011.02893}, 2020.

\bibitem[Yang et~al.(2019)Yang, Swanson, Jin, Coley, Eiden, Gao, Guzman-Perez,
  Hopper, Kelley, Mathea, et~al.]{yang2019analyzing}
Yang, K., Swanson, K., Jin, W., Coley, C., Eiden, P., Gao, H., Guzman-Perez,
  A., Hopper, T., Kelley, B., Mathea, M., et~al.
\newblock Analyzing learned molecular representations for property prediction.
\newblock \emph{Journal of chemical information and modeling}, 59\penalty0
  (8):\penalty0 3370--3388, 2019.

\bibitem[Ying et~al.(2019)Ying, Bourgeois, You, Zitnik, and
  Leskovec]{ying2019gnnexplainer}
Ying, R., Bourgeois, D., You, J., Zitnik, M., and Leskovec, J.
\newblock Gnnexplainer: Generating explanations for graph neural networks.
\newblock \emph{Advances in neural information processing systems},
  32:\penalty0 9240, 2019.

\bibitem[Yu et~al.(2021)Yu, Xu, Rong, Bian, Huang, and He]{yu2021graph}
Yu, J., Xu, T., Rong, Y., Bian, Y., Huang, J., and He, R.
\newblock Graph information bottleneck for subgraph recognition.
\newblock In \emph{International Conference on Learning Representations}, 2021.
\newblock URL \url{https://openreview.net/forum?id=bM4Iqfg8M2k}.

\bibitem[Zhou et~al.(2016)Zhou, Khosla, Lapedriza, Oliva, and
  Torralba]{zhou2016learning}
Zhou, B., Khosla, A., Lapedriza, A., Oliva, A., and Torralba, A.
\newblock Learning deep features for discriminative localization.
\newblock In \emph{Proceedings of the IEEE conference on computer vision and
  pattern recognition}, pp.\  2921--2929, 2016.

\end{thebibliography}
\bibliographystyle{icml2021}



\end{document}